\documentclass[conference]{IEEEtran}
\IEEEoverridecommandlockouts
\usepackage{cite}
\usepackage{amsmath,amssymb,amsfonts}
\usepackage{algorithmic}
\usepackage{graphicx}
\usepackage{textcomp}
\usepackage{xcolor}
\usepackage{float}
\usepackage{hyperref}
\hypersetup{hidelinks,
	colorlinks=true,
	allcolors=black,
	pdfstartview=Fit,
	breaklinks=true}

\def\BibTeX{{\rm B\kern-.05em{\sc i\kern-.025em b}\kern-.08em
    T\kern-.1667em\lower.7ex\hbox{E}\kern-.125emX}}
\begin{document}

\title{Convolutional Recurrent Neural Network with Attention for 3D Speech Enhancement 
}

\author{
\IEEEauthorblockN{\normalsize {Han Yin, Jisheng Bai, Mou Wang, Siwei Huang, Yafei Jia, Jianfeng Chen}}
\IEEEauthorblockN{\normalsize{Joint Laboratory of Environmental Sound Sensing, School of Marine Science and Technology,} \\
\normalsize{Northwestern Polytechnical University, Xi’an, China}\\
\normalsize{E-mail:$\{$yinhan, baijs, wangmou21, hsw838866721, jyf2020260709$\}$@mail.nwpu.edu.cn, chenjf@nwpu.edu.cn}
}}

\maketitle

\begin{abstract}
3D speech enhancement can effectively improve the auditory experience and plays a crucial role in augmented reality technology.
However, traditional convolutional-based speech enhancement methods have limitations in extracting dynamic voice information.
In this paper, we incorporate a dual-path recurrent neural network block into the U-Net to iteratively extract dynamic audio information in both the time and frequency domains.
And an attention mechanism is proposed to fuse the original signal, reference signal, and generated masks.
Moreover, we introduce a loss function to simultaneously optimize the network in the time-frequency and time domains.
Experimental results show that our system outperforms the state-of-the-art systems on the dataset of ICASSP L3DAS23 challenge.
\end{abstract}

\section{Introduction}
3D audio is a multi-channel audio processing technology designed to provide a more immersive listening experience by simulating a realistic three-dimensional sound field.
But in real world environments, speech communication is usually disturbed by various background noise, leading to poor intelligibility and clarity. 
Some speech-related tasks, such as automatic speech recognition, also suffer from performance degradation when noise and reverberation are involved. 
Therefore, speech enhancement (SE), which aims at improving the quality of speech signals, has received widespread attention recently.

Researchers often use filters with different characteristics to eliminate noise components\cite{b1,b2,b3}. 
Another approach named spectral subtraction uses mathematical models to estimate the noise spectrogram, which is then be subtracted from the original spectrogram\cite{b4,b5}. 
However, these conventional algorithms are usually based on strong assumptions and require professional knowledge to model and analyze signals.

To overcome the limitations of traditional algorithms, deep learning (DL) has emerged as a powerful approach to improve the performance of SE\cite{b6,b7,b8}.
It does not require extensive feature engineering or domain knowledge to capture data patterns accurately and can adapt well to unstructured data. 
Recently, SE is typically regarded as a supervised learning problem with neural networks falling into two categories: time-domain-based\cite{b10} and time-frequency (T-F) domain-based methods\cite{b12,b13}. 
Time-domain-based approaches extract information directly from the waveform to construct a regression function for the target speech. 
However, SE methods in the time domain usually need to deal with continuous time series data, which leads to the problem of high computational complexity.
In contrast, SE systems carrying out in the T-F domain can take advantage of efficient frequency-domain algorithms such as Fast Fourier Transform, so the computational complexity is relatively lower.
Firstly, speech is transformed into a time-frequency representation by the Short-time Fourier Transform (STFT). 
Then, DL models, such as convolutional neural networks (CNNs), are used to reconstruct the clean speech spectrogram.
Finally, the inverse Short-time Fourier Transform (iSTFT) is applied to generate time domain signals.

Recently, the U-Net structure has been proposed to improve SE performance\cite{b14}. 
It uses skip connections to connect the output of some layers in the encoder with the input of corresponding layers in the decoder. In this way, the decoder can access lower-level feature maps, which preserve more spatial information.
However, this structure is not suitable for modeling long-range interactions, or interactions with variable-length dependencies, which are common to certain forms of audio such as speech\cite{c1}.
Some researchers try to incorporate recurrent neural networks (RNNs) into CNNs to extract dynamic voice information. E.g., the famous music separation model, Demucs\cite{b15}, adds a bidirectional long short-term memory (Bi-LSTM) block between the encoder and the decoder, achieving state-of-the-art performance. 
Similarly, DCCRN\cite{b16} incorporates a complex-valued LSTM block into the U-Net structure and achieves great SE performance.

\begin{figure*}[htbp]
\centerline{\includegraphics[width=1.0\textwidth]{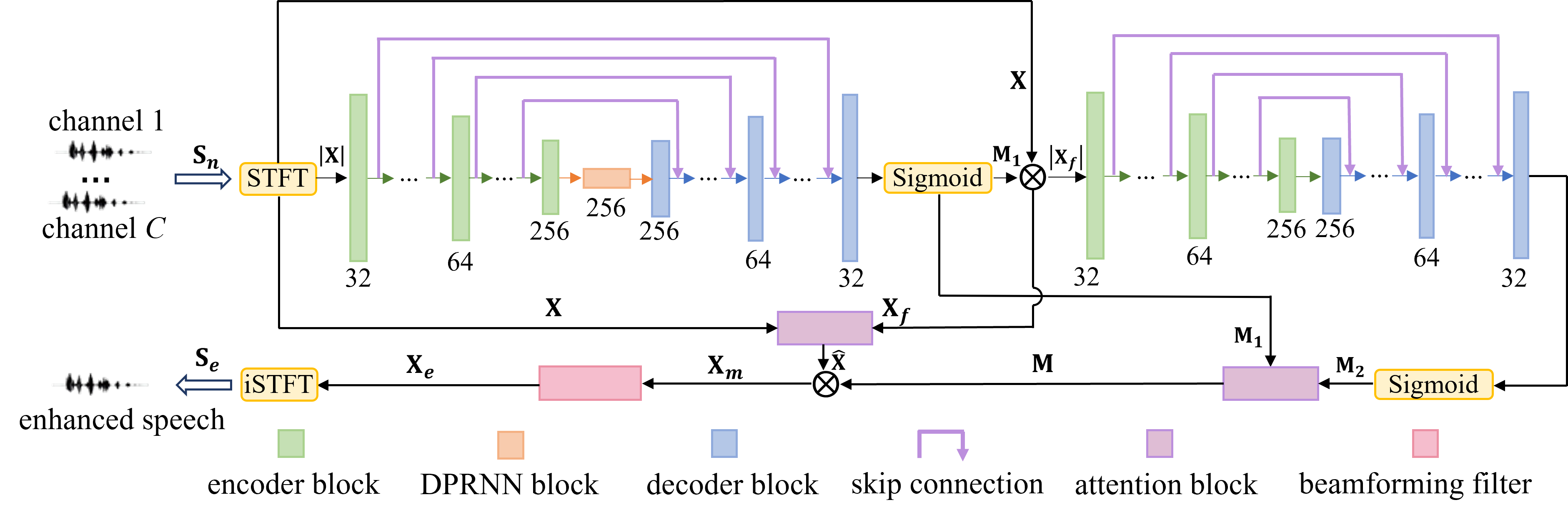}}
\caption{The architecture of the proposed convolutional recurrent neural network for 3D speech enhancement}
\label{fig:system}
\end{figure*}

In addition to one-step approaches, researchers have proposed some multi-stage SE methods, hoping to iteratively eliminate the noise signal\cite{d1,d2,d3,d4}.
In our previous work\cite{c2}, we propose a two-stage U-Net framework to eliminate noise and reverberation in 3D speech signals, which achieves the state-of-the-art performance but lacks some form of attention mechanism for efficient fusion of signals and masks from different stages.

To address the above issues, in this paper, we propose a convolutional recurrent neural network for 3D SE by using two sequentially stacked U-Nets. 
Firstly, we integrate a Dual-path RNN (DPRNN) block\cite{b17} into the first U-Net, which can iteratively and alternately apply time-domain and frequency-domain modeling.
Secondly, we propose an attention mechanism to fuse the original signal, reference signal, and generated masks. 
Then, we utilize the multi-layer perceptron (MLP) to perform frequency-domain beamforming on the multi-channel data to form a monaural output.
Finally, we introduce a loss function to simultaneously optimize the network in the T-F and time domains.

This paper is organized as follows: Section II describes the proposed methods. Section III gives details of experiments such as datasets, evaluation metrics, baselines and settings. Section IV shows the experimental results and Section V gives our discussion and conclusions.

\section{Proposed Methods}
\subsection{Overview}\label{AA}

Fig. \ref{fig:system} illustrates the architecture of the proposed convolutional recurrent neural network for 3D SE. We use the time-domain representation of the noisy speech signal $\mathbf{S}_{n}(t) \in \mathbb{R}^{C \times (T \times S)}$ as the input of the system, where $C$ denotes the total number of audio channels, $T$ represents the duration of the audio, and $S$ is the sampling rate. STFT is then applied to obtain the time-frequency domain representation $\mathbf{X}(t,f) \in \mathbb{C}^{C \times L \times F}$, where $L$ represents the number of frames, and $F$ denotes the frequency bins. 
The convolutional recurrent neural network takes the amplitude of $\mathbf{X}(t,f)$ as input and uses it to reconstruct the spectrogram of clean speech.

$|\mathbf{X}(t,f)|$ is passed through the first U-Net to produce an estimated real-valued mask $\mathbf{M}_1(t,f)\in \mathbb{R}^{C \times L \times F}$. Then, $\mathbf{X}(t,f)$ and $\mathbf{M}_1(t,f)$ are multiplied by element-wise to produce the reference signal $\mathbf{X}_f(t,f) \in \mathbb{C}^{C \times L \times F}$, formulated as:
\begin{equation}
    \mathbf{M}_{1}(t, f)={\rm{UN}}_1[|\mathbf{X}(t, f)|]
\end{equation}
\begin{equation}
    \mathbf{X}_{f}(t, f)=\mathbf{X}(t, f) \odot \mathbf{M}_{1}(t, f)
\end{equation}
Where ${\rm{UN}}_1$ represents the first U-Net, and $\odot$ means element-wise multiplication.

The amplitude of the reference signal is fed into the second U-Net to generate another mask $\mathbf{M}_2(t,f)\in \mathbb{R}^{C \times L \times F}$:
\begin{equation}
    \mathbf{M}_2(t,f) = {\rm{UN}}_2[|\mathbf{X}_f(t, f)|]
    \label{auto2-1}
\end{equation}
Where ${\rm{UN}}_2$ represents the second U-Net.

After that, we use an attention block to perform weighted fusion of mask tensors $\mathbf{M}_1(t,f)$ and $\mathbf{M}_2(t,f)$ to get the final mask $\mathbf{M}(t,f)\in \mathbb{R}^{C \times L \times F}$.
Similarly, original signal $\mathbf{X}(t, f)$ and the reference signal $\mathbf{X}_{f}(t, f)$ are weighted and fused to produce the estimated signal $\hat{\mathbf{X}}(t,f) \in \mathbb{C}^{C \times L \times F}$, formulated as:
\begin{equation}
    \left\{\begin{array}{c}\mathbf{M}(t, f)={\rm{Att}}\left[\mathbf{M}_{1}(t, f), \mathbf{M}_{2}(t, f)\right] \\ \hat{\mathbf{X}}(t, f)= {\rm{Att}}\left[\mathbf{X}(t, f), \mathbf{X}_{f}(t, f)\right]\end{array}\right.
    \label{atten}
\end{equation}
Where $\rm{Att}$ represents the attention block.

The enhanced spectrogram $\mathbf{X}_{m}(t,f) \in \mathbb{C}^{C \times L \times F}$ can be calculated by element-wise multiplication of the estimated signal $\hat{\mathbf{X}}(t,f)$ and the final mask $\mathbf{M}(t,f)$:
\begin{equation}
    \mathbf{X}_m(t, f)=\hat{\mathbf{X}}(t, f) \odot \mathbf{M}(t, f)
    \label{final1}
\end{equation}

$\mathbf{X}_{m}(t,f)$ is passed through a neural beamforming filter in the frequency domain to get a monaural output $\mathbf{X}_e(t,f) \in \mathbb{C}^{L \times F}$ and we use iSTFT to generate the enhanced time domain speech signal $\mathbf{S}_e(t)\in \mathbb{R}^{T \times S}$, formulated as:
\begin{equation}
    \left\{\begin{array}{c}\mathbf{X}_e(t,f)={\rm{Beam}}\left[\mathbf{X}_{m}(t,f)\right] \\ 
    \mathbf{S}_e(t)= {\rm{iSTFT}}\left[\mathbf{X}_e(t,f)\right]\end{array}\right.
    \label{final2}
\end{equation}

\subsection{U-Net structure}

As shown in Fig. \ref{fig:system}, the proposed system is mainly composed of two U-Nets.

\textbf{Encoder}: The encoder comprises $L=10$ stacked encoder blocks, each of which is composed of a two-dimensional convolution with $C_{in}$ input channels and $C_{out}$ output channels, followed by a batch normalization and a LeakyReLU activation function\cite{b18}.

\textbf{Decoder}: The decoder is the inverse of the encoder and comprises 10 decoder blocks, each of which contains a two-dimensional transposed convolution, followed by a batch normalization and a LeakyReLU activation function. Encoder and decoder parameters are set to the same.

\textbf{Skip connection}: Similar to the multi-channel U-Net structure\cite{b14}, skip connections are used between encoder blocks and corresponding decoder blocks. These skip connections allow the input to be directly transmitted to the output and preserve more feature details.

\begin{figure}[htbp]
\centerline{\includegraphics[width=0.35\textwidth]{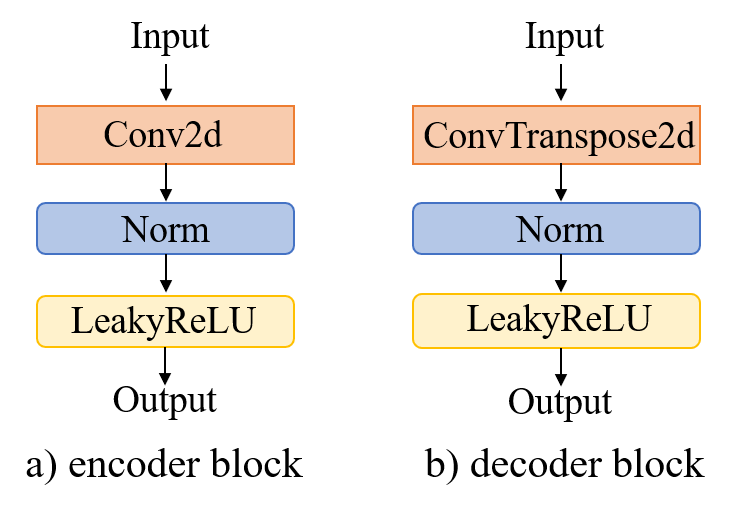}}
\caption{The architecture of the encoder and decoder block}
\label{fig:encoder_decoder}
\end{figure}

The architectures of the encoder and decoder blocks are illustrated in Fig. \ref{fig:encoder_decoder}.
Table \ref{config:conv} presents the detailed configurations of two-dimensional convolution layers in the encoder blocks.
\begin{table}[htbp]
\renewcommand{\arraystretch}{1.2} 
\setlength{\tabcolsep}{12pt}       
\caption{The configurations of Convolution layers in the encoder blocks}
\begin{center}
\scalebox{1.0}{
\begin{tabular}{cccc}
\hline
{$\boldsymbol{L}$} & $\boldsymbol{C_{in}/C_{out}}$ & \textbf{kernel} & \textbf{stride} \\
\hline
1  & $C$/32            & (7,1)  & (1,1)  \\
2  & 32/32            & (1,7)  & (1,1)  \\
3  & 32/32            & (8,6)  & (2,2)  \\
4  & 32/64            & (7,6)  & (1,1)  \\
5  & 64/64            & (6,5)  & (2,2)  \\
6  & 64/96            & (5,5)  & (1,1)  \\
7  & 96/96            & (6,3)  & (2,2)  \\
8  & 96/96            & (5,3)  & (1,1)  \\
9  & 96/128           & (6,3)  & (2,1)  \\
10 & 128/256           & (5,3)  & (1,1)  \\
\hline
\end{tabular}
}
\label{config:conv}
\end{center}
\end{table}

\subsection{DPRNN Block}

In order to perform dynamic feature extraction,
we incorporate a DPRNN block between the encoder and decoder to iteratively apply time-domain and frequency-domain modeling.

\begin{figure}[htbp]
\centerline{\includegraphics[width=0.4\textwidth]{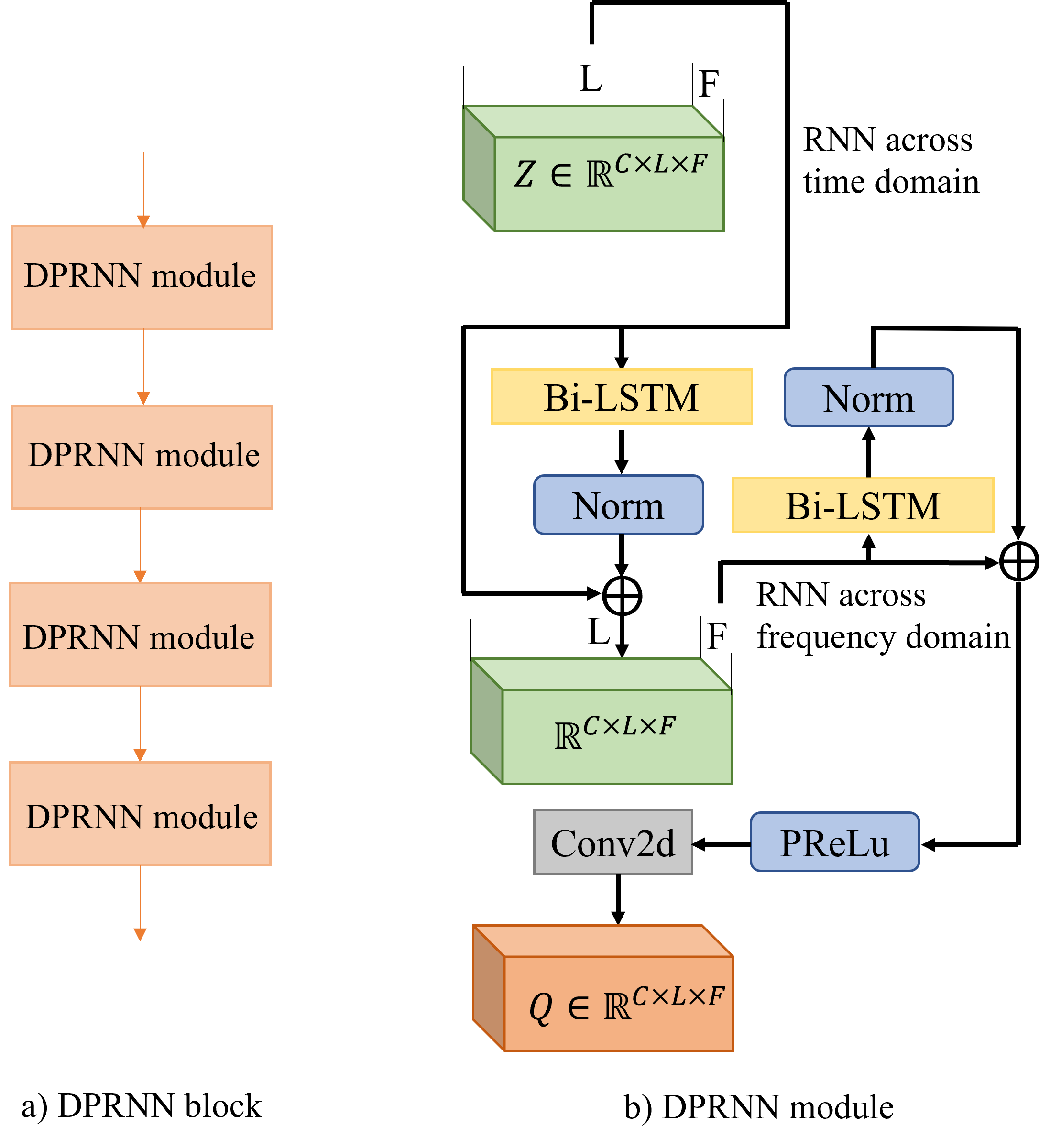}}
\caption{The architecture of DPRNN block and DPRNN module}
\label{fig:DPRNN}
\end{figure}

As shown in Fig. \ref{fig:DPRNN}, the DPRNN block is composed of four stacked DPRNN modules. 
Suppose $\mathbf{Z}(t,f) \in \mathbb{R}^{C \times L \times F}$  is the output of the encoder, which is used as the input to the DPRNN block. 
Subsequently, $\mathbf{Z}$ is passed through two consecutive Bi-LSTMs, each of which contains 128 hidden units, for sequentially time and frequency modeling.
We perform a residual connection between the input and the output of each Bi-LSTM.
After that, the output of the second Bi-LSTM is fed to a convolutional layer with $C_{in}$ of 128 and $C_{out}$ of 256. 
The Norm and PReLU depicted in the figure refer to group normalization\cite{b19} and parametric rectified linear unit\cite{b20}, respectively.

\subsection{Attention Block}
Let $\mathbf{X}_{i}$ and $\mathbf{A}_i$ denote the $i$-th input tensor and $i$-th learnable weight tensor, respectively. The output of the attention block is given by:
\begin{equation}
\mathbf{O}=\sum_{i=1}^{N} \mathbf{A}_i \odot \mathbf{X}_{i}
\label{attention}
\end{equation}

By adding attention blocks to the system, the convolutional recurrent neural network can learn to assign bigger weights to more important features.
Reference signals estimated at each stage can be effectively fused through the attention mechanism to further improve the performance of clean speech spectrogram reconstruction.

\subsection{Neural Beamforming Filter}
\begin{figure}[htbp]
\centerline{\includegraphics[width=0.5\textwidth]{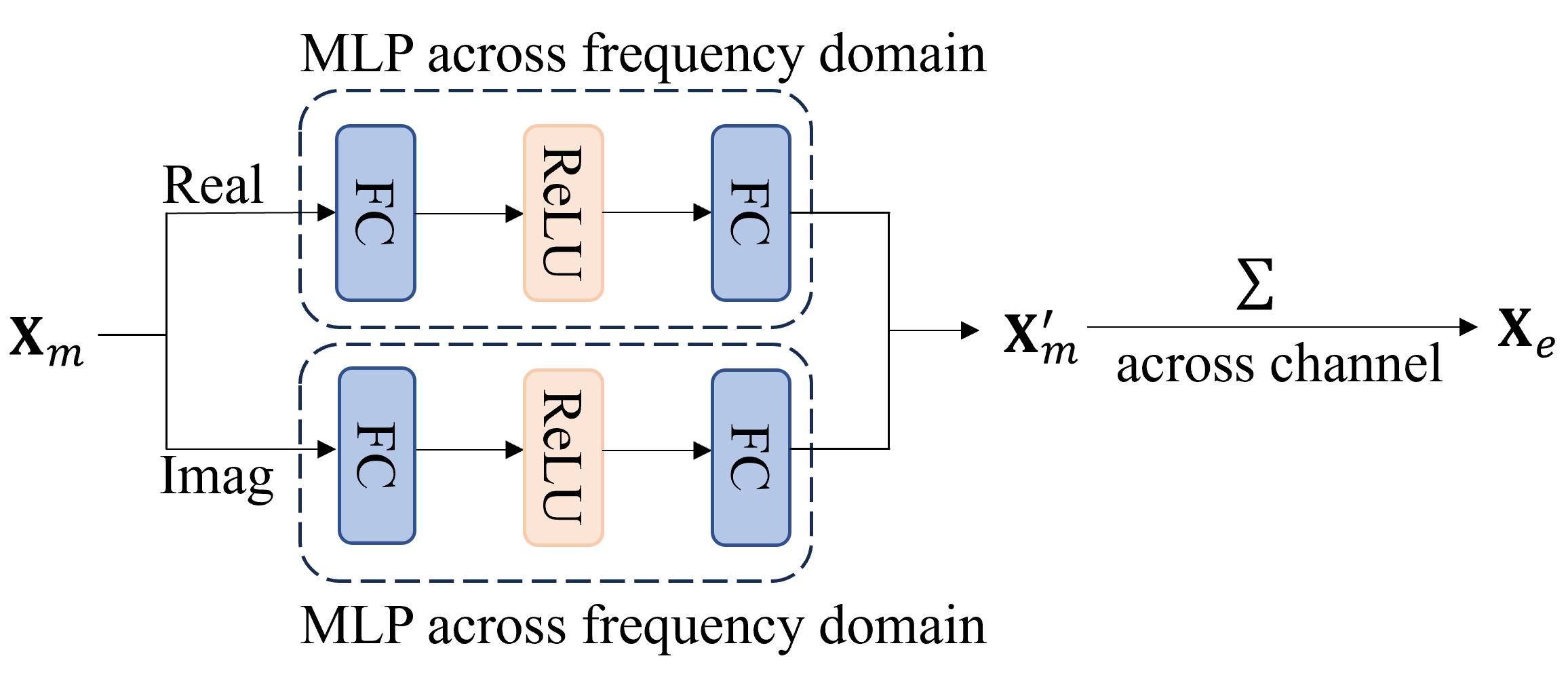}}
\caption{The architecture of the beamforming filter}
\label{fig:beam}
\end{figure}

As shown in Fig. \ref{fig:system}, we pass the enhanced multi-channel spectrogram $\mathbf{X}_{m}(t,f) \in \mathbb{C}^{C \times L \times F}$ through a neural beamformer at the end of the system. There are two motivations: 
1) Through frequency-domain beamforming, the amplitude and phase of $\mathbf{X}_{m}$ at different frequencies are adjusted to enhance the signal from the target direction. 
2) Form a monaural output.
Specifically, as shown in Fig. \ref{fig:beam} , we implement this neural beamforming filter using MLPs so that the beam weights can be learned adaptively.

\subsection{Loss Function}

Recently, the mean absolute error loss function, defined in the time domain, has been increasingly employed for speech enhancement tasks because of its superior performance\cite{b14}, which can be formulated as:
\begin{equation}
\text{MAE} = \frac{1}{T} \sum_{t=1}^{T} |x(t) - \hat{x}(t)|
\end{equation}
where $x(t)$ denotes the clean speech signal and $\hat{x}(t)$ denotes the enhanced speech signal.

In this paper, we propose a loss function which combines the normalized relative error in the T-F and time domains. It can be formulated as:
\begin{equation}
\begin{aligned}
    Loss =&\frac{\gamma}{L F} \sum_{t=1}^{L} \sum_{f=1}^{F}\frac{||\mathbf{X}(t, f)|-| \widehat{\mathbf{X}}(t, f)||}{|\mathbf{X}(t, f)|} \\
        &+\frac{1-\gamma}{T} \sum_{t=1}^{T} \frac{|x(t)-\hat{x}(t)|}{|x(t)|}
    \label{T-Floss}
\end{aligned}
\end{equation}
where $\mathbf{X}(t,f)$ and $\widehat{\mathbf{X}}(t,f)$ denote the clean and the enhanced spectrogram, respectively, $x(t)$ and $\hat{x}(t)$ represent corresponding speech signals in the time domain and $\gamma$ is the hyperparameter.
By default, $\gamma$ is set to 0.5.

\section{Experiments}
\subsection{Dataset}

The proposed system is evaluated on the 3D audio dataset provided by the L3DAS23 challenge.
The dataset is semi-synthetic, where the spatial sound scenes are generated by convolving computed room impulse responses (RIRs) with clean sound samples. The noise data is sourced from the FSD50K dataset\cite{b21}, and clean speech samples are extracted from Librispeech\cite{b22} . Overall, the dataset comprises approximately 90 hours 4-channel recordings with a sampling rate of 16kHz.

To capture a comprehensive range of acoustic environments, two Ambisonics microphones are positioned at 443 random locations across 68 houses to generate RIRs. The sound sources are placed in random locations of a cylindrical grid. One microphone is located in the exact position selected, while the other is positioned 20 cm apart from it. Both microphones are situated at a height of 1.6 m, which approximates the average ear height of a standing person. Details about the dataset can be found at \href{https://www.l3das.com/icassp2023/dataset.html}{https://www.l3das.com/icassp2023/dataset.html}. 

\subsection{Baselines}

We conduct a comparative analysis of the proposed method against several state-of-the-art baselines, including:

\textbf{FasNet}\cite{b23}: A neural beamformer that leverages both magnitude and phase information of the signal, operating in the time domain. 

\textbf{U-Net}\cite{b14}: A multi-channel autoencoder neural network that performs in the T-F domain. This method won the first place in the L3DAS21 challenge, showing its effectiveness in dealing with 3D speech enhancement tasks.

\textbf{SE-UNet}\cite{c2}: A two-stage U-Net architecture proposed by us before, which got the second place in the L3DAS23 challenge. 

\subsection{Evaluation Metrics}

We utilize a set of objective criteria including Short-Time Objective Intelligibility (STOI) and Word Error Rate (WER). STOI is employed to estimate the intelligibility of speech, while WER is used to evaluate the performance of speech on recognition. Both STOI and WER scores are normalized to the range of $[0,1]$, facilitating a consistent and comparative analysis across different methods.

Moreover, we use an additional metric defined by the L3DAS23 challenge, which provides an overall evaluation of speech enhancement performance. This metric combines the STOI and WER scores, which can be calculated as:
\begin{equation}
    \rm{Metric} =\left(\mathrm{STOI}+\left(1-\mathrm{WER}\right)\right) / 2
\end{equation}

\subsection{Training Setup and Hyper-parameters}

We employ a batch size of 12 and the Adam optimizer to train all models. The initial learning rate is set to 0.001, and we use early stopping based on validation loss to select the best model.

For the proposed system, we preprocess audio data using STFT with a window size of 512 and a hop size of 128. We cut all recordings into 4.792-second segments and use them  as the input of the network. Dropout rate is set to 0.1.

\section{Results and Discussion}
 Experimental results of the proposed system and baselines are presented in Table \ref{model results}. 
Results show that the proposed method outperforms baselines in terms of both STOI and WER. 
Compared to our previous work, SE-UNet, STOI is improved by 0.022 and WER is reduced by 0.019.
\begin{table}[htbp]
\renewcommand{\arraystretch}{1.2} 
\setlength{\tabcolsep}{15pt}       
\caption{Experimental results of different systems}
\begin{center}
\scalebox{1.0}{
\begin{tabular}{cccc}
\hline
 \textbf{Model}   & \textbf{STOI}        & \textbf{WER}      & \textbf{Metric}       \\
\hline
FasNet & 0.624       & 0.599      & 0.513       \\
U-Net        & 0.679       & 0.562      & 0.559         \\
SE-UNet          & 0.837       & 0.167      & 0.835         \\
Proposed      & \textbf{0.859}      &\textbf{0.148}     & \textbf{0.856} \\
\hline
\end{tabular}
}
\label{model results}
\end{center}
\end{table}

\begin{figure}[htbp]
\centerline{\includegraphics[width=0.53\textwidth]{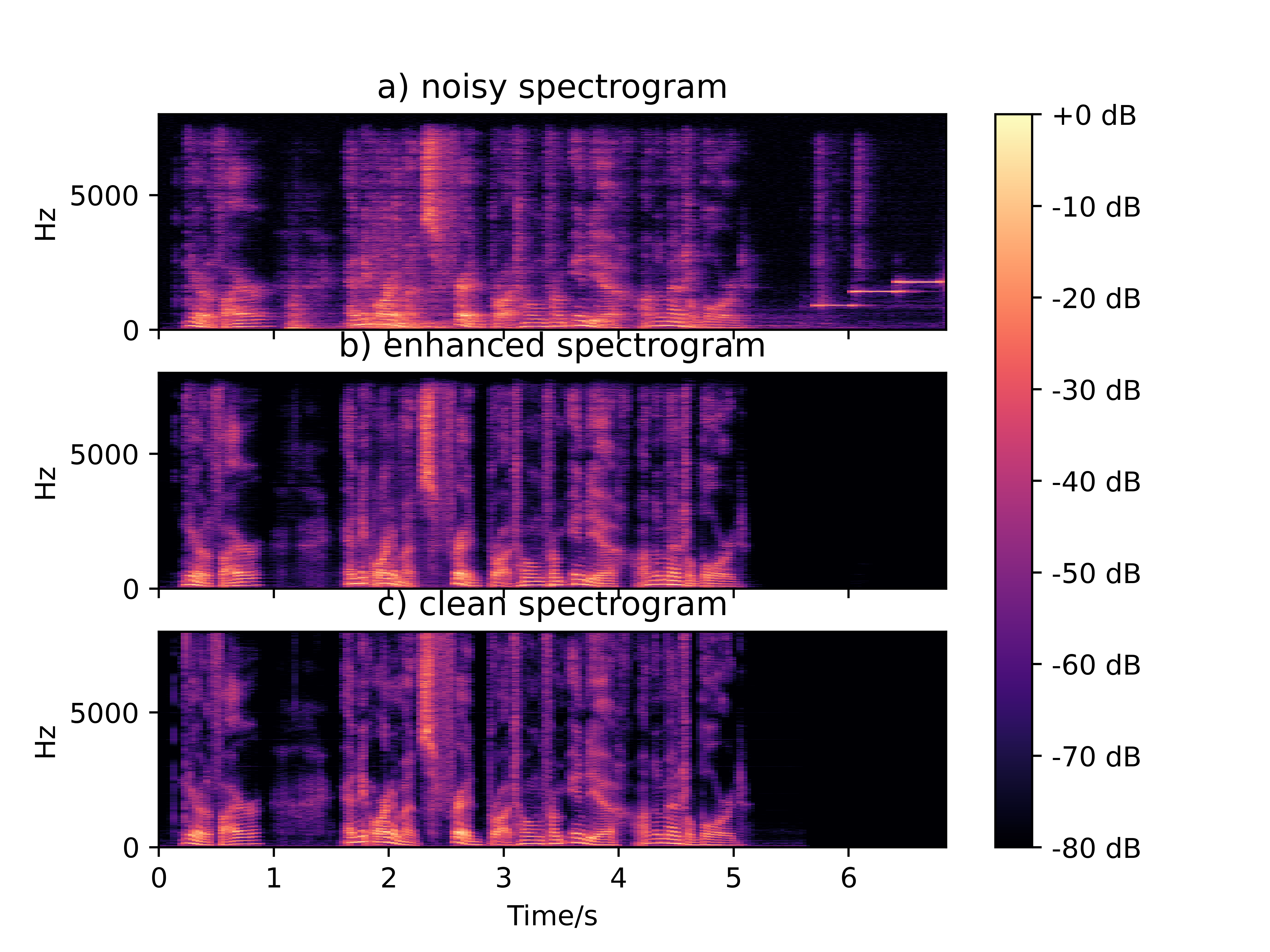}}
\caption{the log-frequency spectrogram of: (a) noisy speech (b) enhanced speech (c) clean speech}
\label{fig:output}
\end{figure}

In Fig. \ref{fig:output}, we visually demonstrate the effectiveness of the proposed method by comparing the enhanced speech spectrogram, the noisy speech spectrogram and the clean speech spectrogram. It can be seen that our system removes the noise signals while suppressing reverberation effectively.

Furthermore, we explore the impact of the proposed loss function by evaluating the system using different settings. As shown in Table \ref{loss_table}, results show that SE performs better when optimized from T-F domain and time domain at the same time. Moreover, it can be found that the T-F loss has a great influence on STOI, while the time-domain loss affects WER more. We recommend to set the default value of $\gamma$ to 0.5 for achieve a general good performance for 3D SE.
\begin{table}[htbp]
\renewcommand{\arraystretch}{1.2} 
\setlength{\tabcolsep}{12pt}       
\caption{Experimental results of different loss functions}
\begin{center}
\scalebox{1.0}{ 
\begin{tabular}{cccc}
\hline
 \textbf{Loss Function}   & \textbf{STOI}        & \textbf{WER}      & \textbf{Metric}       \\
\hline
Proposed($\gamma=0$) & 0.802       & 0.171      & 0.816       \\
Proposed($\gamma=0.5$)       & 0.859     &0.148   & \textbf{0.856} \\
Proposed($\gamma=0.7$)         & \textbf{0.862}      &0.161      & 0.851 \\
Proposed($\gamma=0.3$)       & 0.841      & \textbf{0.143}    & 0.849 \\
\hline
\end{tabular}
}
\label{loss_table}
\end{center}
\end{table}

\section{Conclusions}
In this paper, we propose a convolutional recurrent neural network for 3D speech enhancement using two stacked U-Nets. Firstly, we incorporate a DPRNN block into the first U-Net for alternately extracting dynamic voice information in the time and frequency domains. Then, we introduce an attention mechanism to effectively fuse the original signal, reference signal, and generated masks. Finally, an loss function for SE tasks is proposed to simultaneously optimize the network on the T-F and time domains.
Experimental results show that the proposed system can achieve the state-of-the-art 3D speech enhancement performance and surpasses the baselines on the dataset of L3DAS23 challenge.


\begin{thebibliography}{00}
\bibitem{b1}Doclo S, Moonen M. GSVD-based optimal filtering for single and multimicrophone speech enhancement[J]. IEEE Transactions on signal processing, 2002, 50(9): 2230-2244.
\bibitem{b2}McAulay R, Malpass M. Speech enhancement using a soft-decision noise suppression filter[J]. IEEE Transactions on Acoustics, Speech, and Signal Processing, 1980, 28(2): 137-145.
\bibitem{b3}Paliwal K, Basu A. A speech enhancement method based on Kalman filtering[C]//ICASSP'87. IEEE International Conference on Acoustics, Speech, and Signal Processing. IEEE, 1987, 12: 177-180.
\bibitem{b4}Miyazaki R, Saruwatari H, Inoue T, et al. Musical-noise-free speech enhancement based on optimized iterative spectral subtraction[J]. IEEE Transactions on Audio, Speech, and Language Processing, 2012, 20(7): 2080-2094.
\bibitem{b5}Hasan M K, Salahuddin S, Khan M R. A modified a priori SNR for speech enhancement using spectral subtraction rules[J]. IEEE signal processing letters, 2004, 11(4): 450-453.
\bibitem{b6}Xu Y, Du J, Dai L R, et al. An experimental study on speech enhancement based on deep neural networks[J]. IEEE Signal processing letters, 2013, 21(1): 65-68.
\bibitem{b7}Lu X, Tsao Y, Matsuda S, et al. Speech enhancement based on deep denoising autoencoder[C]//Interspeech. 2013, 2013: 436-440.
\bibitem{b8}Sun L, Du J, Dai L R, et al. Multiple-target deep learning for LSTM-RNN based speech enhancement[C]//2017 Hands-free Speech Communications and Microphone Arrays (HSCMA). IEEE, 2017: 136-140.
\bibitem{b10}Pandey A, Wang D L. A new framework for CNN-based speech enhancement in the time domain[J]. IEEE/ACM Transactions on Audio, Speech, and Language Processing, 2019, 27(7): 1179-1188.
\bibitem{b12}Yuan W. A time–frequency smoothing neural network for speech enhancement[J]. Speech Communication, 2020, 124: 75-84.
\bibitem{b13}Zhao Y, Wang D L. Noisy-Reverberant Speech Enhancement Using DenseUNet with Time-Frequency Attention[C]//Interspeech. 2020, 2020: 3261-3265.
\bibitem{b14}Ren X, Chen L, Zheng X, et al. A neural beamforming network for b-format 3d speech enhancement and recognition[C]//2021 IEEE 31st International Workshop on Machine Learning for Signal Processing (MLSP). IEEE, 2021: 1-6.
\bibitem{c1}Computational analysis of sound scenes and events[M]. Berlin, Germany: Springer International Publishing, 2018.
\bibitem{b15}Défossez A, Usunier N, Bottou L, et al. Music source separation in the waveform domain[J]. arXiv preprint arXiv:1911.13254, 2019.
\bibitem{b16}Hu Y, Liu Y, Lv S, et al. DCCRN: Deep complex convolution recurrent network for phase-aware speech enhancement[J]. arXiv preprint arXiv:2008.00264, 2020.
\bibitem{d1}Wu M, Wang D L. A two-stage algorithm for one-microphone reverberant speech enhancement[J]. IEEE Transactions on Audio, Speech, and Language Processing, 2006, 14(3): 774-784.
\bibitem{d2}Kim J, Hahn M. Speech enhancement using a two-stage network for an efficient boosting strategy[J]. IEEE Signal Processing Letters, 2019, 26(5): 770-774.
\bibitem{d3}Lin J, van Wijngaarden A J L, Wang K C, et al. Speech enhancement using multi-stage self-attentive temporal convolutional networks[J]. IEEE/ACM Transactions on Audio, Speech, and Language Processing, 2021, 29: 3440-3450.
\bibitem{d4}Li J, Luo D, Liu Y, et al. Densely connected multi-stage model with channel wise subband feature for real-time speech enhancement[C]//ICASSP 2021-2021 IEEE International Conference on Acoustics, Speech and Signal Processing (ICASSP). IEEE, 2021: 6638-6642..
\bibitem{c2}Bai J, Huang S, Yin H, et al. 3D Audio Signal Processing Systems for Speech Enhancement and Sound Localization and Detection[C]//ICASSP 2023-2023 IEEE International Conference on Acoustics, Speech and Signal Processing (ICASSP). IEEE, 2023: 1-2.
\bibitem{b17}Luo Y, Chen Z, Yoshioka T. Dual-path rnn: efficient long sequence modeling for time-domain single-channel speech separation[C]//ICASSP 2020-2020 IEEE International Conference on Acoustics, Speech and Signal Processing (ICASSP). IEEE, 2020: 46-50.
\bibitem{b18}Xu J, Li Z, Du B, et al. Reluplex made more practical: Leaky ReLU[C]//2020 IEEE Symposium on Computers and communications (ISCC). IEEE, 2020: 1-7.
\bibitem{b19}Wu Y, He K. Group normalization[C]//Proceedings of the European conference on computer vision (ECCV). 2018: 3-19.
\bibitem{b20}He K, Zhang X, Ren S, et al. Delving deep into rectifiers: Surpassing human-level performance on imagenet classification[C]//Proceedings of the IEEE international conference on computer vision. 2015: 1026-1034.
\bibitem{b21}Fonseca E, Favory X, Pons J, et al. Fsd50k: an open dataset of human-labeled sound events[J]. IEEE/ACM Transactions on Audio, Speech, and Language Processing, 2021, 30: 829-852.
\bibitem{b22}Panayotov V, Chen G, Povey D, et al. Librispeech: an asr corpus based on public domain audio books[C]//2015 IEEE international conference on acoustics, speech and signal processing (ICASSP). IEEE, 2015: 5206-5210.
\bibitem{b23}Luo Y, Han C, Mesgarani N, et al. FaSNet: Low-latency adaptive beamforming for multi-microphone audio processing[C]//2019 IEEE automatic speech recognition and understanding workshop (ASRU). IEEE, 2019: 260-267.
\end{thebibliography}
\end{document}